\documentclass[12pt]{iopart}

\usepackage{iopams}
\usepackage{graphicx}
\usepackage{float}
\usepackage{hyperref}

\begin{document}

\title{Recent advances on integrated quantum communications}

\author{Adeline Orieux and Eleni Diamanti}
\date{}

\address{LTCI, CNRS, T\'el\'ecom ParisTech, Universit\'e Paris-Saclay, 75013, Paris, France}
\ead{adeline.orieux@telecom-paristech.fr, eleni.diamanti@telecom-paristech.fr}
\vspace{10pt}
\begin{indented}
\item[]June 2016
\end{indented}

\begin{abstract}
In recent years, the use of integrated technologies for applications in the field of quantum information processing and communications has made great progress. The resulting devices feature valuable characteristics such as scalability, reproducibility, low cost and interconnectivity, and have the potential to revolutionize our computation and communication practices in the future, much in the way that electronic integrated circuits have drastically transformed our information processing capacities since the last century. Among the multiple applications of integrated quantum technologies, this review will focus on typical components of quantum communication systems and on overall integrated system operation characteristics. We are interested in particular in the use of photonic integration platforms for developing devices necessary in quantum communications, including sources, detectors and both passive and active optical elements. We also illustrate the challenges associated with performing quantum communications on chip, by using the case study of quantum key distribution - the most advanced application of quantum information science. We conclude with promising perspectives in this field.
\end{abstract}


\section{Introduction}
\label{sec:intro}

Quantum information science aims at harnessing quantum mechanical effects to develop systems that can provide a dramatic improvement in information processing, communication efficiency and security. Quantum technologies enable, for example, the distribution of secret keys to two distant parties with unconditional security, which is impossible by classical means \cite{SBC:rmp09}, while the basic principles of quantum computing models have successfully been demonstrated in small-scale systems \cite{BKB:science12}. The true disruptive potential of quantum information technologies, however, can be fully exploited in large-scale systems, namely computing servers manipulating and storing thousands of bits of quantum information (qubits) or networks linking thousands of users performing communication and distributed computation tasks. Similarly to the silicon chip revolution which is at the origin of modern communications, scaling up quantum systems will naturally necessitate the use of integrated technologies. Indeed, chip-based quantum information holds a lot of promise for future high performance information and communication infrastructures, but also presents a lot of challenges as it encompasses several different disciplines. For these reasons, it is one of the most active fields of research today.

Current integration efforts concern all aspects of quantum information, including in particular quantum computing and quantum communications, each of which puts specific constraints on the underlying physical technology. In this review, we will focus on quantum communication technologies on chip (with typical chip dimensions ranging from the millimeter to the centimeter scale).
In this setting, photons are the ideal information carriers due to their long coherence time, their weak interaction with the environment and their high speed. Additionally, and crucially, photons provide a viable pathway to integration, which is necessary to overcome the difficulties inherent in bulk systems to maintain mechanical stability in apparatuses of increasing size and complexity. In principle, integrated photonics can bring together many desirable characteristics in terms of efficiency, cost, scalability, flexibility and performance required for quantum communications.

In the following, we will first review the prevailing integration platforms used for quantum communications, with a particular emphasis on specific features and criteria that determine their suitability for specific applications. We will then describe efforts towards the integration of the main functionalities required in quantum communication systems, namely the generation of nonclassical states of light, their manipulation using reconfigurable passive circuits or active elements for modulation, routing and switching, their storage using quantum memories, and their detection using single-photon or coherent detectors. Beyond the integration of individual components, we are particularly interested here in the on-chip system design aspect, which plays a central role for practical applications. From this perspective, we will discuss in some detail integration efforts for quantum key distribution as well as progress towards on-chip quantum teleportation, entanglement distribution and quantum repeaters, which constitute the core of large-scale quantum networks. We will conclude with the multiple remaining challenges in this field and perspectives for the next few years.

We remark that this review is by no means exhaustive. Integrated quantum communications is a rapidly evolving research field with continuous developments; our goal here is to provide a comprehensive view of the current state of the art with the major tasks accomplished so far and those to be addressed next.

\section{Integration platforms for quantum communications}
\label{sec:platforms}

Thanks to the importance of miniaturization for practical and scalable information and communication devices, today we have at our disposal a diverse set of integration platforms allowing for high performance elements of increasing complexity. The suitability of these platforms for quantum communications is in general evaluated by various important features, including for example the necessity to use foundry services and compatibility with mass manufacturing processes, the ability to support nonlinear and electro-optic effects and single-photon detection, the compatibility with specific encodings of quantum information, such as polarization, path, and time bins, and the adaptivity to a practical communication network infrastructure. The prevailing platforms for such applications can be described as follows:

\begin{itemize}
\item Silicon-based platforms \cite{VP}, which include silicon (Si) as well as silicon nitride (SiN) and silicon carbide (SiC), provide popular solutions as they combine several appealing characteristics. The global prevalence of the Si integration platform for electronics applications has ensured that all the physical and technological attributes of this platform have been fully examined. The existing silicon-based CMOS (Complementary Metal-Oxide-Semiconductor) wafer fabrication facilities open up vast possibilities for cost-effective solutions featuring high yield and reproducibility. Indeed, silicon-on-insulator (SOI) photonics has become a key technology for applications in telecommunications, optical interconnects, medical screening, spectroscopy, biology and chemical sensing, which were inconceivable a few years ago. This technology offers in general waveguides with a very high refractive index, allowing for an increased circuit compactness even for complex layouts involving many elements. Additionally, the high electric field intensity resulting from the extreme mode confinement is beneficial for nonlinear effects. These favorable characteristics, however, come at the expense of poor mode-matching with optical fibers, which is the privileged channel for quantum communications, and increased propagation losses.

\item Platforms based on III-V compound semiconductors \cite{AdachiIIIV2005}, including indium phosphide (InP), gallium arsenide (GaAs), and gallium nitride (GaN), are widely used in optoelectronics as their direct bandgap allows laser emission and gives more favorable conditions than the indirect gap of Si in terms of speed. In the telecommunication wavelengths range, compounds such as InAsP allow laser emission, others such as AlGaAs are transparent and can be used to propagate and manipulate photons, while InGaAs can be used for detection. The III-V materials have a high refractive index and, in addition, they generally present a very strong second-order nonlinearity, allowing for compact and efficient frequency conversion devices or parametric down-conversion sources. These platforms also require foundry services, which are in general less advanced than the CMOS facilities, but are nonetheless widely developed for the laser diodes market. Full system integration, including electrically pumped lasers typically used in applications, may be envisaged with this technology. As for the Si platform, mode-matching with the optical fibers and propagation losses are an issue.

\item Nonlinear optical dielectric materials, in particular lithium niobate (LiNbO$_3$) and potassium titanyl phosphate (commonly known as KTP), form a particularly versatile integration platform. These materials present strong second-order nonlinearities and electro-optic properties, which makes them ideal for parametric down-conversion, frequency conversion and modulation processes. Their fabrication is well studied and controlled, however because of their relatively large size these platforms are less favorable for the development of scalable systems.They have been used for quantum communications with immense success leading to unprecedented developments in the field in the last decade (for a relatively recent review see \cite{Tanzilli:LPR2012}).

\item Platforms producing glass waveguides, such as silica-on-silicon (where silica is the common name of silicon dioxide, SiO$_2$) \cite{PCR:science08} or femtosecond laser writing \cite{fslaserwriting:1996,Osellame:JOSAB2003}, have been developed and used for quantum communications more recently, leading to reduced propagation losses and excellent mode-matching with standard optical fibers. These are also versatile platforms, with a well controlled fabrication procedure that does not require time-consuming foundry services and makes them ideal for rapid device and system tests. These platforms cannot support functionalities requiring nonlinear or electro-optic effects but they are well adapted for complex linear circuits involving a large number of integrated beam splitters (known as directional couplers). Contrary to the Si and III-V platforms, femtosecond laser written technologies also offer the capacity to preserve and manipulate polarization easily, which is important for some applications in quantum communications, and enable a remarkable versatility in the 3-dimensional geometry of the circuits.
\end{itemize}

In addition to the above integration platforms that have been widely used for quantum communications, other important integrated technologies include nitrogen vacancy centers in diamond \cite{HBD:nature15}, which can be coupled to superconducting circuits \cite{Esteve:PRL107}, and semiconductor quantum dots interfaced with photonic nanostructures \cite{LMS:rmp15}. These platforms are being developed primarily for applications in quantum computing and quantum simulation, which fall beyond the scope of this review; however, some results obtained with these technologies also have implications in quantum communications and quantum networks \cite{HBD:nature15,KBK:pnas15}.

It is clear from the above descriptions that no single integration platform can gather all the desired characteristics for a specific application. Indeed, it is generally understood that the next generation of quantum communication devices and systems will adopt hybrid integration technologies with the goal of bringing together the best elements of each platform. In the following, we will describe some important achievements obtained with the aforementioned technologies and discuss future directions towards advanced quantum communication networks.

\section{Integrated quantum communication devices}
\label{sec:devices}

The main optical quantum communication devices aim at generating, manipulating, storing and detecting quantum states of light. One important design element for all devices is the operation wavelength; communication through optical fibers imposes the use of the telecommunication wavelength range, where propagation loss is minimum, while free-space communications typically require near-infrared wavelengths. In the following, we describe integration efforts in all types of devices employed in quantum communication systems.

\subsection{Generation}
\label{subsec:generation}

The first key element of an optical quantum communication system is a source of nonclassical states of light. Some quantum communication applications, such as standard quantum key distribution (QKD) protocols \cite{SBC:rmp09}, only require single-photon states. As discussed in detail in recent reviews \cite{EFM:rsi11,SZ:jmo12}, such states can be generated either in a deterministic manner using single-photon emitters or probabilistically by heralding the generation of the desired state in one of the two members of a correlated photon pair with the detection of the second member; finally, single-photon states can also be simply simulated by highly attenuating coherent states produced by lasers. Weak coherent pulses are in fact sufficient for most QKD implementations, but the suppression of multi-photon events achieved by heralded single-photon sources is useful for other quantum communication protocols, such as quantum teleportation. The ideal statistics reached by on-demand single-photon sources, on the other hand, are necessary for applications in quantum computing and quantum simulation, as well as for repeater-based quantum networks. For such networks \cite{NB:natphoton14}, as well as for performing advanced applications, including device-independent \cite{BCP:rmp14} and distributed verification \cite{PCW:prl12} protocols, it is also necessary to generate two or multi-photon entangled states. Indeed, entangled-photon states constitute the most important resource in quantum communications. The performance of entangled-photon sources can be evaluated using several quality factors, for instance, the visibility or fidelity of the generated states with respect to the target state, and the violation of suitable Bell inequalities quantified by the Bell parameter. The degree of freedom where entanglement is generated is also an important characteristic depending on the desired application.

Sources generating correlated photon pairs are the basis both of the heralded generation of single-photon states used in quantum communications and of the generation of two-photon entangled states (also known as Einstein-Podolsky-Rosen - EPR - pairs). Hence, the vast majority of works on integrated sources so far has aimed at the generation of such states. Currently, the best on-chip sources of photon pairs at telecom wavelengths make use of second-order and third-order nonlinear processes, in particular, spontaneous parametric down-conversion (SPDC) and spontaneous four-wave mixing (SFWM), respectively. The former is possible in the III-V platform as well as in periodically poled LiNbO$_3$ (PPLN) and KTP (PPKTP) waveguides, while only the latter can be used in the Si platform.

\begin{figure}[htb]
\centerline{\includegraphics[width=\textwidth]{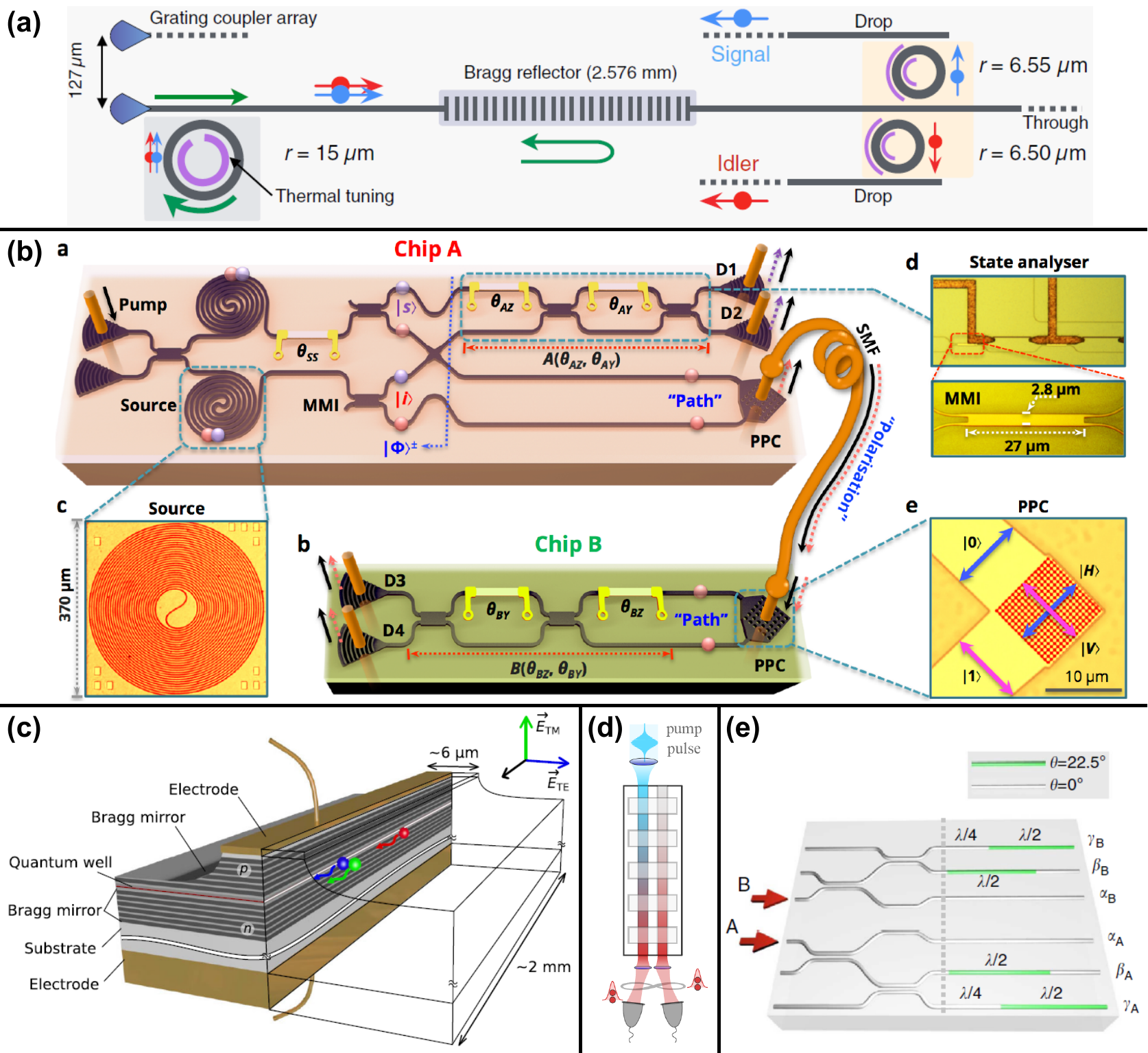}} \caption{Examples of integrated devices for the generation and manipulation of quantum photonic states. (a) Integrated silicon chip with a tunable microring SFWM source and tunable filters, as well as a pump rejection Bragg filter \cite{Bajoni:PRX2014}. (b) Silicon chips with fiber interconnects, spiralled SFWM sources of photon pairs, tunable Mach-Zehnder interferometers and path-polarization qubit converters \cite{OBrien:arxiv1508}. (c) Electrically injected SPDC source of photon pairs in an AlGaAs laser diode \cite{Ducci:PRL112}. (d) SPDC source of two-photon NOON states based on two coupled PPLN waveguides \cite{Silberhorn:1505}. (e) Silica chip with femtosecond laser written waveguides for integrated two-photon polarization tomography, containing directional couplers and waveplates \cite{Sansoni:NatComm5}. Figures adapted with permission from: (a) Ref. \cite{Bajoni:PRX2014}, 2014 under Creative Commons \href{URL}{http://dx.doi.org/10.1103/PhysRevX.4.041047}; (b) Ref. \cite{OBrien:arxiv1508}, \copyright 2016 OSA \href{URL}{http://dx.doi.org/10.1364/OPTICA.3.000407}; (c) Ref. \cite{Ducci:PRL112}, \copyright 2014 APS \href{URL}{http://dx.doi.org/10.1103/PhysRevLett.112.183901}; (d) Ref. \cite{Silberhorn:1505}, \copyright 2015 APS \href{URL}{http://dx.doi.org/10.1103/PhysRevA.92.053841}; (e) Ref. \cite{Sansoni:NatComm5}, \copyright 2014 NPG \href{URL}{doi:10.1038/ncomms5249}.}
\label{fig:Devices1}
\end{figure}

In the Si platform, narrowband ($\leq$ 0.1 nm, instead of 1-100 nm typical of standard sources) entangled photon pairs can be generated in high quality factor microrings evanescently coupled to straight waveguides, as shown for example in Fig.~\ref{fig:Devices1}(a). Time-energy entangled photon pairs with 89\% raw visibility \cite{Bajoni:Optica2} and reconfigurable path-entangled pairs with a Bell parameter of 2.69 \cite{OBrien:NatComm6} have been reported with this geometry (we recall that the upper bound of the Bell parameter is $2\sqrt{2}\approx 2.83$ for an ideal state \cite{Cir:lmp80}). Furthermore, compact spiralled waveguides (see Fig.~\ref{fig:Devices1}(b)) combined with on-chip two-photon interference showing raw visibilities larger than 96\% have allowed the generation of two-photon NOON states \cite{OBrien:NatPhot8}. Nanowire waveguides combined with on-chip polarization rotators have also permitted the generation of polarization-entangled states with more than 91\% fidelity with respect to a Bell state \cite{Takesue:SciRep2}. All these highly compact sources have been made possible by the very tight bending radius and strong mode confinement that can be achieved in silicon-on-insulator waveguides. Note that these sources are based on SFWM where the pump wavelength is in general very close to the wavelength of the generated photons, which means that notch filters with a very high extinction are needed to eliminate residual pump photons. On the other hand, standard telecom lasers can be used as pump lasers.

In the III-V platform, different phase-matching techniques have led to promising photon pair sources with various quantum state properties in AlGaAs waveguides. The propagation losses usually associated with the quasi-phase-matching technique in AlGaAs have been sufficiently reduced recently, allowing for the generation of high quality entangled photon pairs with a coincidence-to-accidental ratio (CAR) larger than 100 \cite{Qian:APL103,Qian:OL39}. Furthermore, counterpropagating phase-matching, in which the pump beam impinges transversally on the waveguide surface and generates photons that are spatially separated in two countepropagating guided modes \cite{Ducci:PRL110}, enables the generation of intrinsically narrowband ($<$ 1 nm) polarization-entangled pairs. Finally, modal phase-matching, in which a pump mode guided in a photonic bandgap defect mode is converted to photon pairs guided by total internal reflection, has shown broadband emission of photon pairs: polarization-entangled \cite{Helmy:OE21} and energy-time-entangled \cite{Ducci:optica16} pairs have been obtained with similar raw Bell parameters of about 2.6. Note that a first demonstration of an electrically pumped source generating its own pump laser within the nonlinear waveguide (see Fig.~\ref{fig:Devices1}(c)) with a CAR larger than 13 has been demonstrated using this modal phase-matching strategy \cite{Ducci:PRL112}, opening the way towards fully integrated sources.

As mentioned in Section \ref{sec:platforms}, sources based on PPLN or PPKTP waveguides are the most established ones and have shown impressive quality and efficiency for many years. Some recent results include a source of path-entangled photon pairs based on two coupled PPLN waveguides (see Fig.~\ref{fig:Devices1}(d)) with a fidelity larger than 84\% with respect to a two-photon NOON state \cite{Silberhorn:1505}, a source of single-longitudinal-mode pairs with a 60 MHz bandwidth \cite{Silberhorn:NJP17} for applications with atom-based quantum memories, and a source of polarization-entangled photon pairs with a very high raw Bell parameter of 2.82 \cite{Tanzilli:OComm327}. A source of heralded single photons with a heralding efficiency of 60\%, one of the best reported so far for integrated sources, has been achieved in a LiNbO$_3$ chip combining a PPLN waveguide for SPDC and a wavelength demultiplexer for separating the emitted photons \cite{Silberhorn:NJP15}.

In silica waveguides, the nonlinear effect is not strong enough to allow the generation of photon pairs on reasonably small sizes. Indeed, SFWM in silica generally requires waveguide lengths of several tens of centimeters and up to a few meters, typically only achieved in optical fibers. An envisioned strategy in this case is the hybrid integration of nonlinear dielectric crystal waveguides for the source part combined with glass chips suitable for photon manipulation (discussed in the following section). While this promising direction is pursued, some recent works have combined off-chip devices with such chips to generate quantum states of particular interest. For example, a silica-on-silicon chip consisting of four tunable beam splitters was used to entangle two squeezed beams generated by (off-chip) sub-threshold optical parametric oscillators (OPOs) \cite{OBrien:NatPhot9}. The entanglement performance of this source, which is adapted to quantum information protocols exploiting continuous variables (CV) of light, was mainly limited by coupling losses in and out of the chip. Another illustration of this partially integrated approach was the use of femtosecond laser written glass circuits in combination with SPDC in a BiB$_3$O$_6$ crystal for the generation and characterization of single-photon W states \cite{GHP:natphoton14}.

Some progress has also been made recently towards the generation of multi-photon states required for advanced quantum communication protocols. From a practical application perspective, it is particularly interesting to be able to herald the generation of these states instead of using commonly employed post-selection techniques \cite{BTC:njp13}. Heralded multi-photon states have been generated in experiments using BBO crystals \cite{Walther:NatPhot2010} or PPLN and PPKTP waveguides \cite{Jennewein:NatPhot2014}, while, importantly, the generation of a photon triplet has been reported recently in dielectric waveguides \cite{Silberhorn:QELS}.

We remark that a major issue, common to all the aforementioned sources, is the probabilistic nature and Poissonian statistics of the (multi-)photon state emission that leads to very small generation rates limited by undesired higher photon number terms. Techniques using multiplexing of several heralded sources combined with delay lines and active switches offer a promising route towards the solution of this problem \cite{Eggleton:NatComm4}.

\subsection{Manipulation}
\label{subsec:manipulation}

In all quantum information applications, the nonclassical states used as the main resources for the computation or communication protocols need to be suitably manipulated by implementing specific operations. For optical systems, the linear optics quantum computing (LOQC) scheme introduced in 2001 by Knill, Laflamme and Milburn (KLM) \cite{KLM} and widely studied since then, showed that the only basic elements required in addition to single-photon sources and detectors are beam splitters and phase-shifters, when the qubits are encoded in the path of the photons (also known as dual-rail encoding). The same scheme can be used with polarization-encoded photonic qubits by using wave-plates and polarizing beam splitters instead \cite{KLMpolar}.

The vast majority of the quantum photonics circuits demonstrated to date have used the dual-rail encoding because the building blocks are much easier to integrate than their polarization counterparts. Significant recent results include a silica-on-silicon reconfigurable circuit of nested Mach-Zehnder interferometers with electrically tunable phase-shifters \cite{OBrien:Science349}, an AlGaAs tunable Mach-Zehnder interferometer relying on the Pockels effect \cite{OBrien:OComm327}, a UV-written glass waveguide circuit for on-chip quantum teleportation with thermo-optic phase-shifters \cite{Walmsley:NatPhot8} (see Fig.~\ref{fig:tel}), femtosecond laser written glass circuits with up to 13 input and output optical modes for Boson Sampling \cite{Sciarrino:SciAdv1} and the previously mentioned femtosecond laser written glass circuits for the creation of single-photon W states over 16 spatial modes \cite{GHP:natphoton14}. Femtosecond laser written glass circuits have also been used recently to manipulate polarization-encoded photons \cite{Sansoni:NatComm5} (see Fig.~\ref{fig:Devices1}(e)) and even polarization-path hyper-entangled photons \cite{Sansoni:NatPhot7,Orieux:LSA2015}.

Note that many of the developed circuits have been designed for visible or near-infrared light because of the better performance of currently available sources and detectors at these wavelengths; however all the demonstrated devices can be adapted straightforwardly for telecom wavelengths by adjusting the waveguide cross-sections, as the materials used for the chips are also transparent in the telecom range. Indeed, thermally reconfigurable femtosecond laser written glass circuits at telecom wavelengths have been demonstrated very recently \cite{FMR:lsa15}.

Additional components are needed to interconnect the chips performing photon manipulation and the optical fibers of the communication network. In particular, on photonic circuits with dual-rail encoding, it is necessary to convert the qubits to an encoding that is more adapted for propagation in optical fibers, such as polarization or time-bin encoding. Such integrated qubit converters have been demonstrated recently on silicon-on-insulator chips (see Fig.~\ref{fig:Devices1}(b)) with 2D grating couplers allowing to interconvert polarization and path qubits with a fidelity of 98\% \cite{OBrien:arxiv1508,Massar:OL38}. For semiconductor-based platforms (silicon and III-V alike), because of the large refractive index difference between the optical fibers and the waveguides of the chip, it is also necessary to devise an efficient way of transferring one guided mode to the other. Several strategies can be adopted depending on the encoding: if there is only one input polarization mode then 1D grating couplers can be used to couple light incident from the chip surface into a waveguide, while if polarization matters then tapered waveguides and butt-coupled fibers on the chip edge should be used instead \cite{VP}. Spectral filters are also required inside the chips, especially just after the photon sources, to remove the powerful pump beam so that it does not generate noise in the detectors or cause unwanted nonlinear effects in the circuit. These filters have to be designed carefully so that they have low loss and preserve entanglement. One example of such integrated filters combined with a source of quantum states has been realized in a silicon chip \cite{Bajoni:PRX2014} (see Fig.~\ref{fig:Devices1}(a)) with more than 95 dB of rejection of the pump.

Finally, it may be useful to convert single photons from one wavelength to another in some particular cases, for example, if efficient detectors or quantum memories are available in a wavelength range different from the telecom one. This conversion, to be truly useful, must be done efficiently at the single-photon level and must not degrade the quantum state by noise addition. Such a process has been reported recently with PPLN waveguides to convert single photons from 910 nm to 600 nm with an internal conversion efficiency over 70\% and similar signal-to-noise ratio before and after conversion \cite{NIST:PRL2012}.

\subsection{Storage}
\label{subsec:storage}

A crucial element of future quantum communication networks are quantum memories, namely devices able to store a quantum state for a tunable amount of time before restituting it on-demand with high fidelity (see \cite{Tittel:JMO60,Review:QMemoryApplications} for recent reviews). Indeed, such a device is an essential part of quantum repeaters, allowing to increase the achievable communication distance between two users that is currently limited by the propagation losses of optical fibers. By enabling the synchronization of photonic processes and hence deterministic system operation, quantum memories also provide an efficient way to deal with the probabilistic nature of the multi-photon state emission, and are therefore a key factor for scalability. We note that the benefits of quantum memories extend well beyond this first application, as pointed out in Ref. \cite{Review:QMemoryApplications}. Indeed, they could also provide a way to manipulate the photons, for instance, as spectral shapers or frequency converters.

To date, no integrated memory directly compatible with optical telecommunication networks has been demonstrated. Recently, two integrated memories have shown promising results to store red frequency photons: one of them relies on a LiNbO$_3$ waveguide doped with thulium ions \cite{Tittel:PRL108}, while the other used the femtosecond laser writing technique to fabricate a waveguide inside a Y$_2$SiO$_5$ crystal doped with praseodymium ions \cite{CSM:arxiv15}. A cesium vapour-filled hollow core fiber operated at room temperature \cite{Sprague:NJP15} has also shown encouraging results. Note that these memories would be compatible with telecommunication networks by associating them with photon pair sources emitting one photon in the telecom range and one photon in the visible region of the spectrum where most accessible atomic transitions occur. To avoid the need for such specific dual band photon sources or single-photon frequency conversion devices, telecom compatible memories based on Erbium-doped fibers \cite{Tittel:PRL115} have been investigated recently. Much remains to be done in this domain to find a practical way of integrating these memories and achieve the high fidelities, efficiencies and long storage times needed for applications. A major difficulty encountered is, among others, the bandwidth mismatch between the atomic transitions (a few GHz) and typical photon sources (a few tens of GHz to a few THz).

\subsection{Detection}
\label{subsec:detection}

Finally, a ubiquitous component of quantum communication systems exploiting single-photon properties are single-photon detectors. In bulk experiments performed at telecom wavelengths, InGaAs single-photon avalanche photodiodes (SPADs) are typically used; however their detection efficiency at these wavelengths is generally limited to around 10-20\% as they are plagued with too much dark count noise at higher efficiencies. This is why a lot of effort has been devoted to an alternative approach, namely superconducting single-photon detectors (SSPDs), in which the absorption of single photons induces local heat that can be detected by a change in the resistivity of a superconducting wire (for a relatively recent review see \cite{Review:SNSPD2012}). Indeed, these detectors offer much larger detection efficiencies than SPADs, with very little dark count noise and a reduced jitter. Additionally, and crucially for some quantum information applications, they allow photon-number resolving detection.

\begin{figure}[htb]
\centerline{\includegraphics[width=\textwidth]{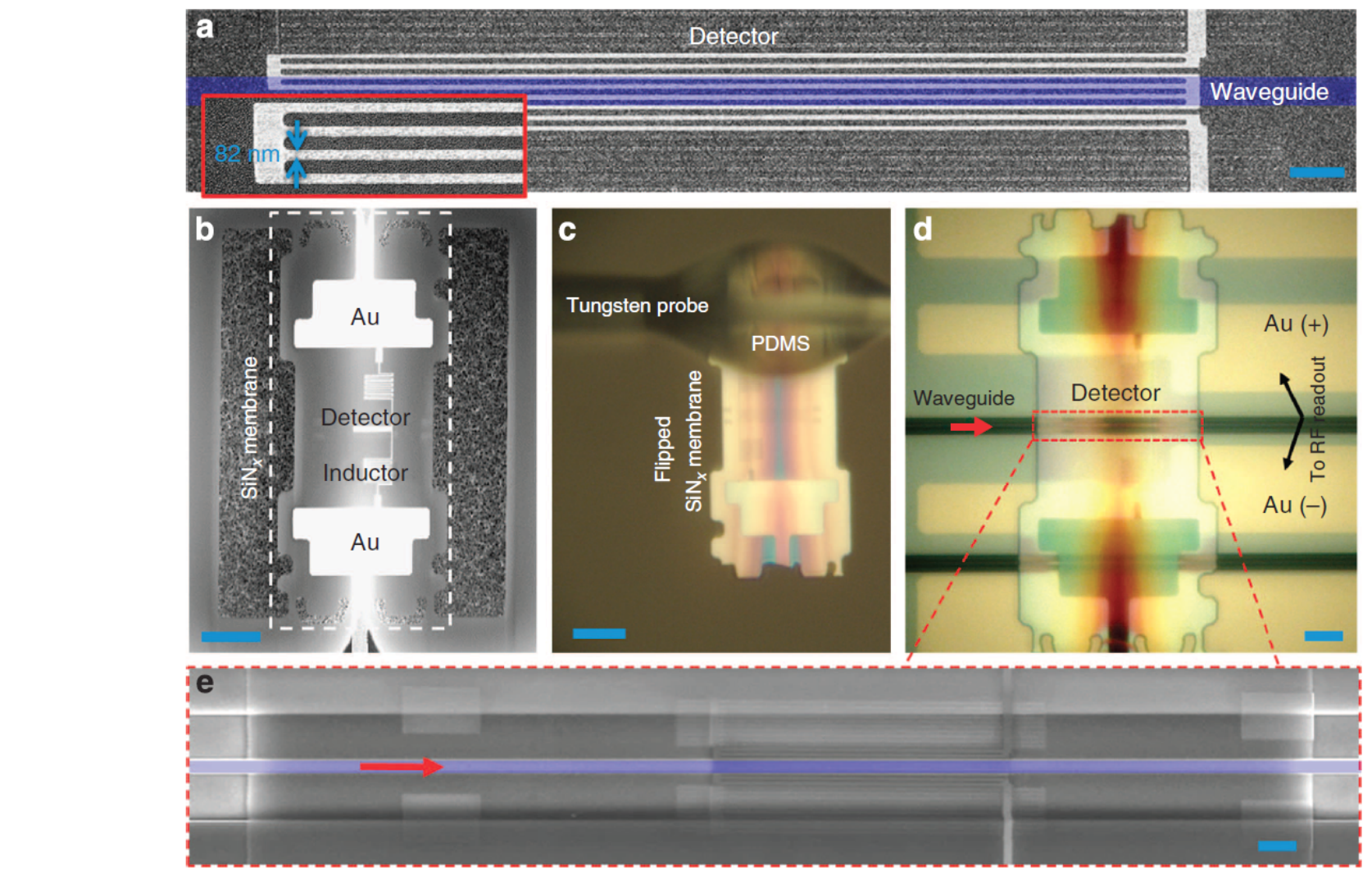}} \caption{NbN SNSPDs integrated on top of AlN waveguides by SiN membrane transfer. The superconducting NbN nanowire detector before transfer can be seen in panel a. Panels b, c and d show the steps of the transfer of the SiN membrane containing the detector onto the AlN waveguide, by means of a tungsten microprobe. Panel e shows the final device with the detector on top of the waveguide. Figure reproduced from Ref. \cite{Englund:NatComm6}, 2015 under Creative Commons \href{URL}{doi:10.1038/ncomms6873}.}
\label{fig:Devices2}
\end{figure}

In general, the integration of single-photon detectors is challenging. SSPDs present some advantages in this respect, as the deposition of a superconducting nanowire on top of a waveguide allows the photons to evanescently couple to the detector and be efficiently absorbed on short propagation lengths (see Fig.~\ref{fig:Devices2}). This technique is in principle compatible with all the different integration platforms we have introduced. The main challenge resides in finding a good fabrication technique such that the outstanding performances of bulk superconducting detectors are preserved in the integrated geometry. Several high quality results have been demonstrated for such superconducting nanowire single-photon detectors (SNSPDs) recently with NbN nanowires deposited on top of SiN \cite{Pernice:SR5}, Si and AlN \cite{Englund:NatComm6} or AlGaAs \cite{Fiore:IEEE2015} waveguides, yielding on-chip detection efficiencies of more than 70\%, 50\% and 20\% respectively. Despite these very promising results, one remaining limitation of these detectors is that they need to be cooled at cryogenic temperatures, which necessarily complicates practical network deployments and may hinder their integration on the same chip with other components performing better at room temperature.

As we will see in the following section, on-chip detection is currently the main obstacle on the way to fully integrated qubit-based quantum communication systems, while for CV-based quantum communications which require coherent detection techniques it is possible to use standard devices developed by the silicon photonics industry.

\section{Integrated quantum communication systems}
\label{sec:systems}

In the previous section we have seen a great number of studies targeting high performance on-chip quantum communication devices. In view of future large-scale quantum networks incorporating such devices, it is also crucial to take a system design approach including the components themselves but also the surrounding network environment and the additional constrains imposed by the implemented quantum communication protocols. Interestingly, although the performance of on-chip components is not necessarily superior to that of their bulk counterparts at the first stages of development, the full advantage of integration may appear once the entire system performance has been characterized.

There is a wide range of protocols and applications pertaining to quantum communications \cite{GT:natphoton07}. By far the most developed is quantum key distribution (QKD), while other important quantum cryptographic primitives include bit commitment, coin flipping, oblivious transfer, secret sharing, digital signatures, anonymous communication, secure identification, etc. Quantum communication complexity protocols, such as quantum fingeprinting, as well as random number generation are also essential elements of quantum communications networks. Finally, the backbone of such networks is entanglement distribution, leading to quantum teleportation, quantum relays and quantum repeaters. In the following, we will review recent integration efforts for QKD and for entanglement distribution and quantum teleportation; these case studies illustrate the challenges linked to system integration and pave the way to on-chip implementation of all other aforementioned protocols and, ultimately, of quantum communication networks.

\subsection{Quantum key distribution}
\label{subsec:qkd}

The ability to distribute secret keys between two parties over an untrusted channel with unconditional (or, information-theoretic) security, that is regardless of the capacities of a malicious eavesdropper, is arguably one of the most powerful achievements of quantum information science. Since the first proposal of a QKD protocol in 1984 \cite{BB84}, this application has advanced tremendously \cite{SBC:rmp09,LCT:natphoton14} and is now leading the way to the industrial development of quantum technologies. Despite its importance, however, efforts towards the development of chip-scale integrated QKD systems have been limited until recently. This development will be crucial for moving on from the current bulky, costly systems to compact and lightweight devices that can be mass manufactured at low cost. In this way, integration can open the way to the wide adoption of quantum technologies for securing communications in quantum information networks.

In QKD implementations, the key information is typically encoded either in discrete variables (DV), such as the polarization or phase of single photons (or more commonly in practical implementations, of weak coherent pulses, as discussed earlier), or in continuous variables (CV), such as the values of the quadrature components of the quantized electromagnetic field, those for instance of coherent states. These states are then transmitted over optical fibers or free space and detected at the receiver's end using single-photon detectors for DV-QKD protocols or coherent (homodyne or heterodyne) detection techniques for CV-QKD protocols. A prominent example of a DV-QKD protocol is decoy-state BB84 \cite{LMC:prl05,Wang:prl05}, while Gaussian modulated coherent state CV-QKD \cite{GG:prl02} is the most commonly used protocol in the CV framework. Finally, the distributed-phase-reference protocols, such as differential phase shift (DPS) QKD \cite{IWY:prl02} and coherent one way (COW) \cite{SBG:apl05}, where the key information is encoded on the phase difference between adjacent weak coherent pulses and on photon arrival times, respectively, also require the use of single-photon detectors. Beyond these standard protocols, there have been major recent advances in QKD, including in particular the proposal of measurement-device-independent (MDI) QKD \cite{LCQ:prl12} (see also \cite{BP:prl12}) and of the so-called Robin-Round (RR) DPS QKD protocol that does not require monitoring the signal disturbance to establish security \cite{SWK:nature14}. The former scheme provides a practical way of eliminating security breaches due to the imperfections of the receiver's detectors, while the latter features a very high noise tolerance. Both schemes have been demonstrated experimentally, for instance, in Refs. \cite{TYZ:prx15,VLC:jmo15,CLF:natphoton16,TST:natphoton15,WYC:natphoton15,GCL:prl15}, but their implementation would greatly benefit from photonic integration; this is especially true for RR-DPS QKD, which relies on a complex setup involving multiple interferometers, time delay circuits, etc.

Because of the birefringence of optical fibers, polarization encoding in DV-QKD is not always desirable, especially for long-distance communications; for this reason, the most advanced systems typically encode the key information on phase or time bins, which brings the need for precise and stable interferometers. These are in fact used in all phase-encoding-based protocols, including DPS QKD for example. They are the first components in QKD systems to have been integrated; indeed, planar lightwave circuits (PLCs) based on silica-on-silicon technology have been used for many years as asymmetric Mach-Zehnder interferometers \cite{DTL:opex06,YFT:ol12} because of their low loss and temperature-stabilized operation.

\begin{figure}[htb]
\centerline{\includegraphics[width=\textwidth]{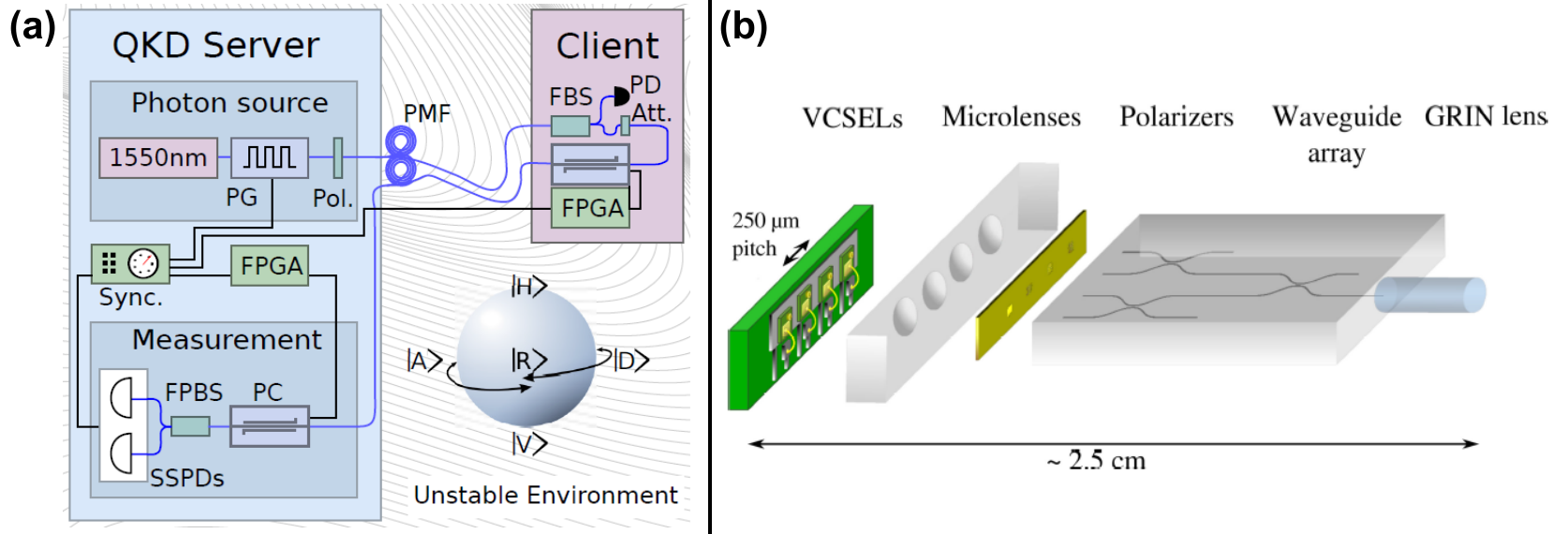}} \caption{(a) Experimental setup for client-server reference frame independent QKD \cite{ZAM:prl14}. Most components are standard fiber-based telecom ones, such as the 1550 nm laser, the pulse generator (PG), and the beam-splitters (FBS) and polarizing beam-splitters (FPBS). The polarization controllers (PC) are integrated on a LiNbO$_3$ chip. (b) Design of a handheld QKD transmitter device \cite{VRF:jstqe15} combining in a single integrated system several components fabricated using different technologies: lasers, lenses, polarizers and a waveguide array. Figures adapted with permission from: (a) Ref. \cite{ZAM:prl14}, \copyright 2014 APS \href{URL}{http://dx.doi.org/10.1103/PhysRevLett.112.130501}; (b) Ref. \cite{VRF:jstqe15}, \copyright 2015 IEEE \href{URL}{10.1109/JSTQE.2014.2364131}.}
\label{fig:qkd1}
\end{figure}

Although the use of PLC-based devices illustrates the favorable characteristics of on-chip QKD system components, significant efforts are required to develop fully integrated systems. Early steps in this direction considered a client-server scenario, where the QKD client (Alice) holds a low cost, lightweight device with integrated elements while the server (Bob) incorporates the large system resources that are difficult to integrate, in particular the single-photon detectors for DV-QKD. This scenario was motivated precisely by this difficulty that we have discussed in Section \ref{subsec:detection}, and could have some applications in use cases where many users receive secret keys by a few providers; MDI QKD, for instance, is well adapted to this scenario. In Ref.~\cite{ZAM:prl14}, the developed miniaturized client device includes a LiNbO$_3$ integrated polarization controller in a system implementing the so-called reference frame independent QKD protocol in a two-way optical fiber configuration where the server includes both the photon source and the single-photon detectors (Fig.~\ref{fig:qkd1}(a)). Ref.~\cite{VRF:jstqe15} demonstrated a handheld device for Alice, tailored for polarization-encoded, short range, free space QKD. It is based on an integrated optics architecture combining various techniques. In particular, vertical cavity surface emitting lasers, coupled to micro-polarizers fabricated using lithography, are used to generate the polarization qubits. They are combined with a waveguide array, fabricated using femtosecond laser writing on glass (which is suitable for polarization encoding as we have seen previously) for ensuring the spatial overlap of these qubits (Fig.~\ref{fig:qkd1}(b)).

\begin{figure}[htb]
\centerline{\includegraphics[width=\textwidth]{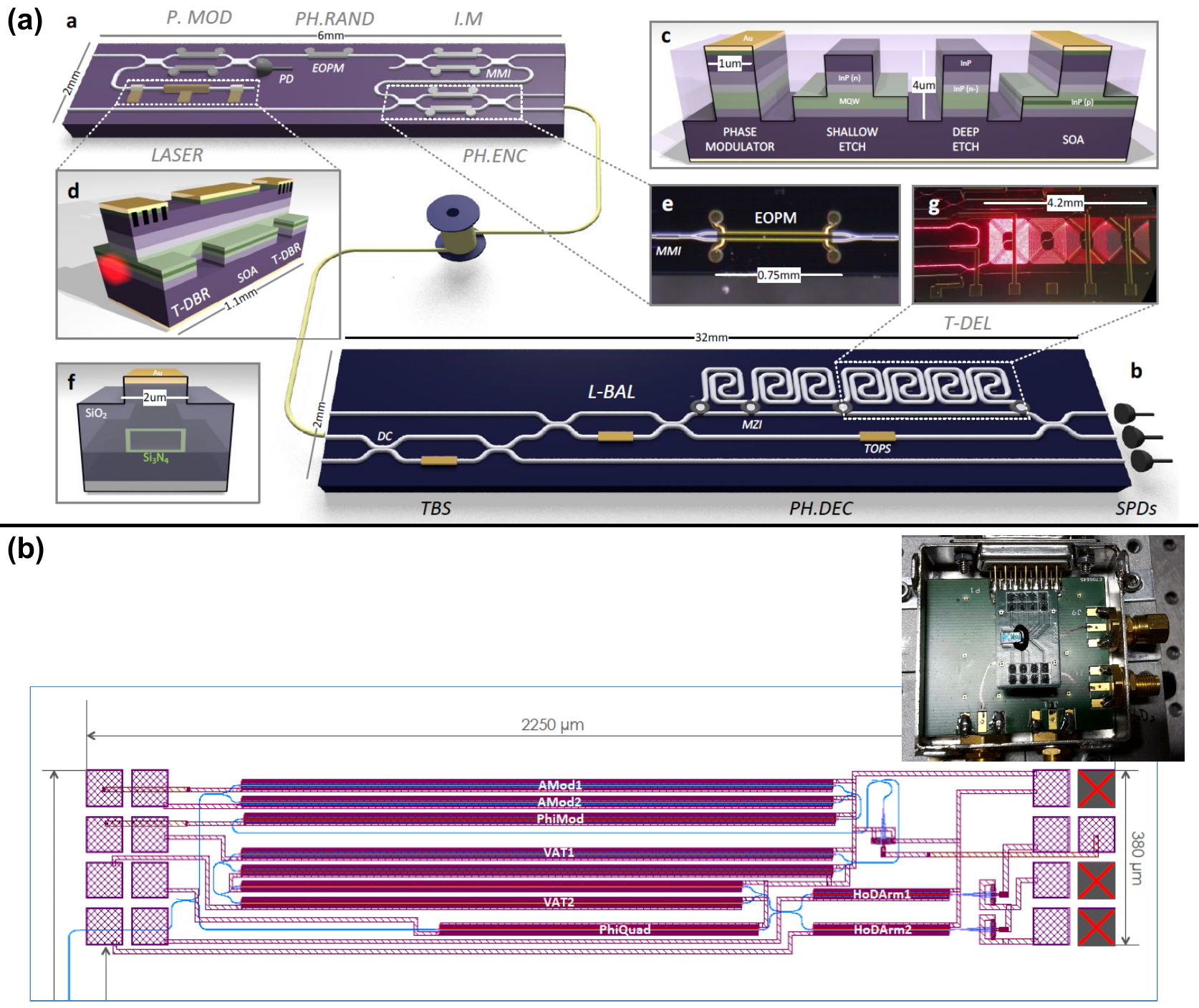}} \caption{(a) Photonic chip architecture combining several integrated devices for the implementation of discrete-variable and distributed-phase-reference QKD protocols \cite{SEG:arxiv15}. The first chip (panel a, with details in panels c, d and e) is built on an InP substrate with active components, including a laser, optical amplifiers and phase modulators. The second chip (panel b, with details in panels f and g) is based on the SiON platform and plays the role of the receiver circuit, with Mach-Zehnder interferometers and delay lines. (b) Silicon photonic chip design for continuous-variable QKD. The inset shows a picture of the packaged chip including several such systems and stand-alone integrated components, such as homodyne and heterodyne detectors. Figures adapted with permission from (a) Ref. \cite{SEG:arxiv15}, \copyright 2015 \href{URL}{https://arxiv.org/abs/1509.00768}, courtesy of Mark Thompson; (b) Ref. \cite{PZH:qipc15}.}
\label{fig:qkd2}
\end{figure}

The above systems provided a proof-of-principle characterization of partially integrated QKD systems, however fully chip-based systems are necessary for a wide range of applications, including long-distance secure communications or key sharing in a server-server scenario, for instance between data centers with high security requirements in a cloud network infrastructure. Such systems are also needed for enhanced functionality and integrability in current communication networks. For DV-QKD and distributed-phase-reference protocols, the main limitation in this direction remains for the moment the need for on-chip single-photon detectors. Notwithstanding this element, a recent experiment demonstrated a high degree of system integration \cite{SEG:arxiv15} (Fig.~\ref{fig:qkd2}(a)). In this system, Alice's module is integrated on InP and includes a tunable telecom laser source and both active and passive elements such as electro-optic phase modulators and Mach-Zehnder interferometers enabling the reconfigurable implementation of the transmitter functionalities of the decoy-state BB84, DPS and COW QKD protocols. Bob's module on the other hand is integrated on SiO$_x$N$_y$ and includes passive elements enabling the receiver functionalities of these protocols, with the exception of the detection stage. This system achieved GHz operation with estimated secret key generation rates on the order of a few hundreds of kbit/s over a channel attenuation corresponding to 20 km of standard optical fiber in laboratory conditions.

The implementation of CV-QKD systems requires only standard telecom components \cite{JKL:natphoton13,DL:entropy15}, hence opening the way to complete system integration, including the coherent detection stage. Progress in this direction has been achieved recently using Si photonics, with a proof-of-principle photonic chip including some of the functionalities required by the Gaussian modulation coherent-state CV-QKD protocol \cite{PZH:qipc15} (Fig.~\ref{fig:qkd2}(b)), including amplitude and phase modulation and shot noise limited homodyne detection using germanium (Ge) photodiodes. Further system integration is feasible and particularly relevant in the context of new advances in this field, towards high-speed CV-QKD systems using a locally generated phase reference signal \cite{QLP:prx15,SBC:prx15,HHL:ol15}.

\subsection{Entanglement distribution and quantum teleportation}
\label{subsec:teleportation}

Intrinsic losses in photonic communication channels eventually make it impractical to perform communication tasks over point-to-point links, and impose a network structure. Quantum networks are indeed crucial for increasing the range of quantum communication systems and will be the neuralgic element of the future quantum Internet \cite{Kimble:nature08}. This task is enabled by entanglement distribution, the fundamental building block of quantum teleportation and quantum repeaters. Because of its importance, long-distance entanglement distribution has been thoroughly investigated experimentally with impressive results, for instance over 300 km of optical fiber \cite{IMT:oe13}. Furthermore, quantum teleportation over a 100-km fiber link \cite{TDS:optica15} and entanglement swapping over a 143-km free-space link \cite{HSF:pnas15} were shown recently. Combined with quantum memories, these protocols lead to embryonic demonstrations of quantum repeaters, such as those of Refs. \cite{BCT:natphoton14,SPZ:arxiv15}, which are suitable for fiber-based quantum communication networks.

The aforementioned advanced implementations are all relying on largely explored nonlinear optical materials, namely PPLN and PPKTP, for the generation of entangled or heralded single-photon states, and superconducting nanowire or avalanche photodiode single-photon detectors for fiber optic and free space experiments, respectively. The PLC interferometers routinely used in phase-encoded QKD systems were also employed in such setups while fiber-based quantum memories were used in Ref. \cite{SPZ:arxiv15}. These efforts towards compact systems are promising, however in general full integration of such implementations is a challenging task.

\begin{figure}[htb]
\centerline{\includegraphics[width=\textwidth]{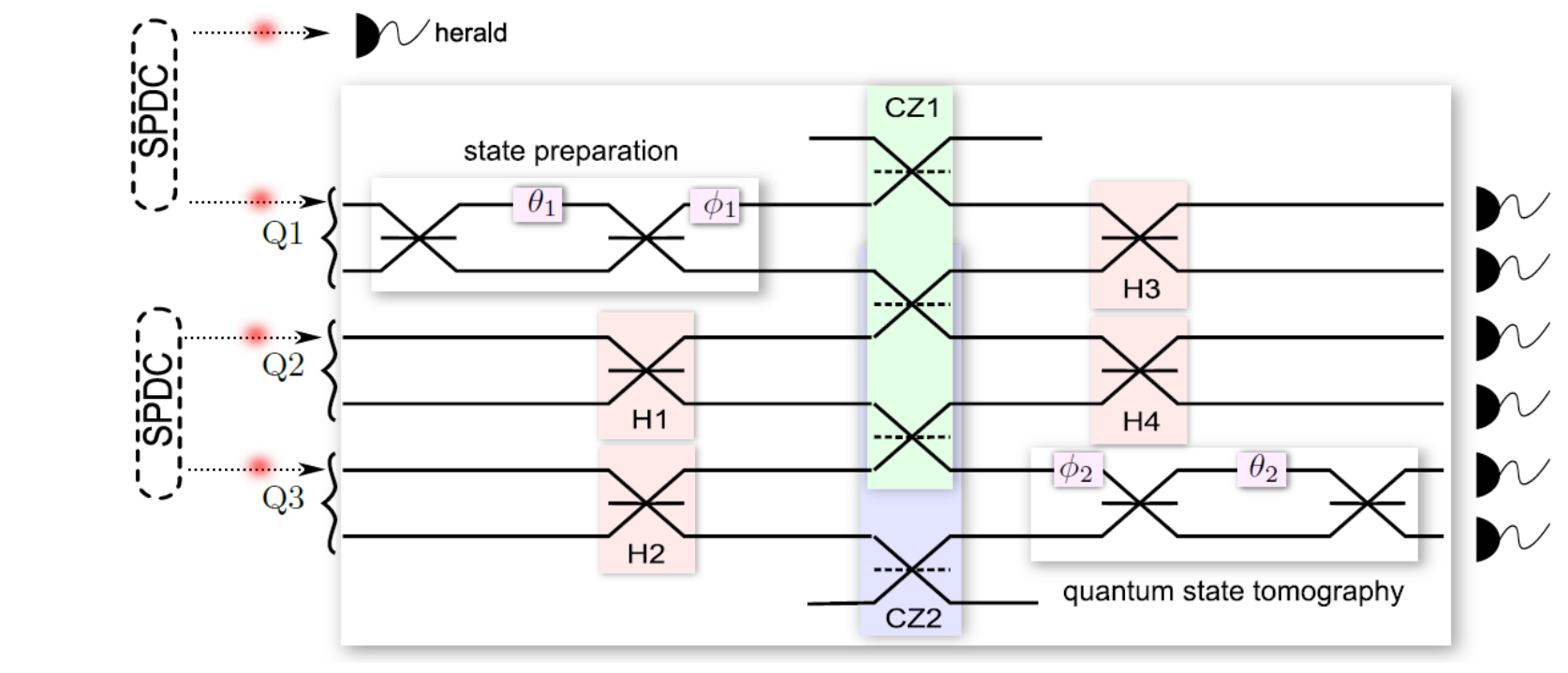}} \caption{Silica-on-silicon chip architecture for quantum teleportation \cite{Walmsley:NatPhot8}. The chip is based on dual-rail encoding and contains several integrated beam-splitters with reflectivities 1/2 or 1/3 and thermo-optics phase-shifters. Figure adapted with permission from Ref. \cite{Walmsley:NatPhot8}, \copyright 2014 NPG \href{URL}{doi:10.1038/nphoton.2014.217}.}
\label{fig:tel}
\end{figure}

Some further steps towards on-chip systems have been taken recently. We remark in particular the quantum teleportation experiment of Ref.~\cite{Walmsley:NatPhot8}, performed using a single UV-written silica-on-silicon chip including directional couplers (integrated beam splitters) and phase shifters for the realization of the path-encoded qubits, Hadamard and controlled-phase gates, and state tomography required by the implemented protocol (Fig.~\ref{fig:tel}). This protocol bypasses the feedforward operation typically required in quantum teleportation, and additionally the parametric down-conversion sources and avalanche photodiodes remain off chip; however, the experimental demonstration of the aforementioned functionalities opens the way to more complex implementations in future integrated systems. As we have previously mentioned, a silica-on-silicon chip was also used for continuous-variable entanglement generation and characterization - using off-chip homodyne detectors \cite{OBrien:NatPhot9}; these are the first steps towards the demonstration of quantum teleportation in the CV framework as well. Finally, Ref.~\cite{OBrien:arxiv1508} showed entanglement distribution, certified by a Bell test violation, between two separate fiber-connected chips using silicon photonics and based on path and polarization encodings (see Fig.~\ref{fig:Devices1}(b)). This is a natural setting in a network, hence such experiments advance towards realistic communication scenarios that rely on flexibility and interconnectivity.

A common feature of the above integrated quantum communication experiments is the absence of any appreciable distance between the communicating parties. Losses are indeed currently a bottleneck in these implementations, and reducing them at all stages of the system (within the chip, fiber coupling, etc) will be of utmost importance in order to improve the system performance and hence develop practical on-chip quantum communication applications.\\

Let us finish this section by remarking that, in addition to developing the integrated quantum communication systems themselves, several other elements come into play in a practical network environment. In particular, the synchronization of the devices typically requires very fast electronics, hence integration of electronic components operating at GHz rates have to be devised. Furthermore, the packaging of the systems, whose operation requires the tuning and routing of a number of interconnected components, needs to satisfy stringent practical constraints. Finally, multiplexing techniques routinely used to increase the bandwidth in data communications and successfully tested in quantum communications, for example, for QKD \cite{PDL:apl14} and entanglement distribution \cite{TGO:jap15}, will need to be adapted to chip-based systems.

\section{Conclusions and outlook}

In this review, we have discussed significant recent advances in the field of integrated quantum communications, which have addressed challenges linked to the main components used in such implementations but also to the system aspects inherent in operation in a network environment. The ingenious solutions that have been devised until today lead to further challenges that need to be tackled next.

On the components side, improving the heralding efficiency of on-chip sources of single and multi-photon states will be crucial, while the development of efficient frequency converters and integrated quantum memories, single-photon detectors and photon counters needs to be pursued using the techniques and integration platforms discussed in the previous sections, potentially adopting hybrid solutions. On the systems side, the realization of the first chip-based QKD and quantum teleportation systems, even at a prototypical stage, is extremely promising. The reduced payload of integrated quantum communication systems opens the way to a great range of applications: for example, mobile QKD networks can be envisaged, while crucially such systems may be deployed on satellites overcoming the challenge of losses, inherent in fiber optic networks, hence bringing quantum communications to the global scale. Further advances in integrated systems, including overcoming the losses and developing multiplexing techniques as discussed previously, will be required to achieve advanced protocol implementations on chip, such as device-independent and measurement-device-independent QKD, distributed communications employing multipartite states, as well as active quantum teleportation and quantum repeater links surpassing the performance of direct transmission links.

It is important to remark that some of the aforementioned protocols have a direct link with tests of fundamental physics. For example, device-independent QKD requires loophole-free Bell tests, which have recently been demonstrated in diamond \cite{HBD:nature15} and using nonlinear waveguides \cite{SMC:arxiv15,GVW:arxiv15}. This type of experiments can become routine thanks to the new opportunities offered by integration. More generally, on-chip systems for long-distance entanglement distribution and quantum communications in space can enable testing the quantum/classical interface and foundational notions such as nonlocality and contextuality in previously inaccessible regimes.

At a practical level, targeting truly useful systems with potential for industrial development will require the further enhancement of available infrastructures, both on the chip fabrication side, by developing worldwide high capacity multiple-use foundry services and growth facilities, and on the network side, by using deployed fibers and satellite devices for the purposes of quantum communication experiments. Furthermore, work towards certification and standardization, which is important for the impact and validation of future applications and already actively pursued for QKD, has to take into account the specificities of chip-based systems as well. Finally, we note that research in integrated quantum communications is of interest to the underlying technologies too; as an example, the requirements of Si-integrated modulator performance for QKD are different from those needed for standard classical optical communications, hence the exploration for suitable characteristics opens up new possibilities more generally in silicon photonics.

Because of the great promise it holds for demonstrating the true disruptive potential of quantum information science in large-scale systems, the field of integrated quantum communications is the subject of extremely active and innovative research work worldwide. Beyond any doubt, the resulting developments in the next years will change the landscape of our future communication and computation capacities and practices.

\ack

ED thanks the participants of the NSF Quantum Information on Chip Workshop held in Padova, Italy, in October 2015, for stimulating discussions; some of the discussed topics are reflected in this review. The authors acknowledge financial support from the City of Paris (project CiQWii), the French National Research Agency (projects QRYPTOS and COMB), the Ile-de-France Region (project QUIN), and the France-USA Partner University Fund (project CRYSP).

\section*{References}

\bibliography{reviewintegratedqc}
\bibliographystyle{iopart-num}

\end{document}